\pdfoutput=1
\documentclass[10pt,aps,prl,nofootinbib,superscriptaddress,twocolumn,preprintnumbers,balancelastpage]{revtex4}

\usepackage{graphicx,epsfig,psfrag,amssymb,hyperref}
\usepackage{multirow}
\usepackage{color,graphicx,epsfig,psfrag,amsmath,empheq}
\usepackage{bm}
\usepackage{mathrsfs,amsfonts,color}
\usepackage{slashed}
\usepackage{hepunits}
\usepackage{color}
\usepackage{enumitem}
\usepackage{xfrac}

\begin{document}

\title{LHCb future dark-sector sensitivity projections for Snowmass 2021}

\author{Daniel Craik}
\email{dancraik@mit.edu}
\affiliation{Laboratory for Nuclear Science, Massachusetts Institute of Technology, Cambridge, MA 02139, U.S.A.}

\author{Phil Ilten}
\email{philten@cern.ch}
\affiliation{Department of Physics, University of Cincinnati, Cincinnati, OH 45221, U.S.A.}

\author{Daniel Johnson}
\email{daniel.johnson@cern.ch}
\affiliation{Laboratory for Nuclear Science, Massachusetts Institute of Technology, Cambridge, MA 02139, U.S.A.}

\author{Mike Williams}
\email{mwill@mit.edu}
\affiliation{Laboratory for Nuclear Science, Massachusetts Institute of Technology, Cambridge, MA 02139, U.S.A.}

\begin{abstract}
We provide future LHCb dark-sector sensitivity projections for use in the Snowmass reports.
These include updated projections for dark photons and the Higgs portal, along with new projections for axion-like particles that couple predominantly to gluons.
\end{abstract}

\maketitle

\section{Introduction}

As part of the Snowmass process, the {\em Rare and Precision Frontier} is coordinating studies of dark sectors and experimental approaches to exploring these paradigms.
The LHCb experiment has already demonstrated world-leading sensitivity to visible dark-photon decays~\cite{LHCb-PAPER-2017-038,LHCb-PAPER-2019-031} and to GeV-scale Higgs-portal scalars~\cite{LHCb-PAPER-2015-036,LHCb-PAPER-2016-052} using Run~2 and Run~1 data, respectively.
In this document, we update existing projections for dark photons and the Higgs portal, and provide new projections for axion-like particles (ALPs) that interact predominantly with the Standard Model (SM) via their coupling to gluons. (We do not address other ALP couplings, such as in Ref.~\cite{CidVidal:2018blh}.)

\textbf{Important note on penguin-based projections}:
Both the Higgs portal and ALP projections are based on penguin decays, which are sensitive to the order of the loop calculations. In addition, the ALP sensitivity depends strongly on the assumptions made about the unknown UV physics. Therefore, comparing our projections to those from other experiments requires some care, as such comparisons are only informative when the same assumptions are made.
Presumably, during the Snowmass process an agreement will be made on the assumptions used to ensure all projections are on an equal footing. Once that is done, we will update this document.
Currently, this document uses the one-loop calculations from Ref.~\cite{Batell:2009jf} and the UV-dependent term taken to be $\log{(\Lambda_{\rm UV}^2/m_t^2)} \pm \mathcal{O}(1) \to 1$ as in Ref.~\cite{Aloni:2018vki}.
This choice produces sensitivities that are an order of magnitude weaker than if we adopted the same penguin calculations and UV scale as Ref.~\cite{Chakraborty:2021wda}.

\section{Dark photons}

The published LHCb dark-photon searches were motivated by Refs.~\cite{Ilten:2015hya,Ilten:2016tkc} which include projections for Run~3 and Run~6.\footnote{Throughout this document, we take the luminosities to be 15/fb in Run~3 and 300/fb in Run~6. We assume future LHCb upgrades maintain the Run~3 performance despite the increased instantaneous luminosity in future runs.}
The LHCb dimuon results from Run~2~\cite{LHCb-PAPER-2017-038,LHCb-PAPER-2019-031} confirm the future dimuon-based projections~\cite{Ilten:2016tkc}. Therefore, there is no need to update the predictions above the dimuon threshold.
For lower masses, Ref.~\cite{Ilten:2015hya} proposed using radiative charm decays as a channel that could be selected online by the LHCb trigger. The primary motivation was that LHCb could not perform electron identification in its first high-level trigger (HLT) stage.
In 2018, LHCb implemented electron identification in the HLT, which permitted selecting inclusive $A' \to e^+e^-$ decays in a similar manner to that in which the published dimuon searches were performed.
Furthermore, the development of the GPU-based {\em Allen} application~\cite{Allen}, which will implement the first HLT stage in Runs~3 and 4, should allow continued use of this inclusive dielectron approach in the future.
Therefore, we do update the LHCb dark-photon sensitivity projections below the dimuon threshold to account for this new capability.

\begin{figure*}[t]
  \centering
  \includegraphics[width=0.99\textwidth]{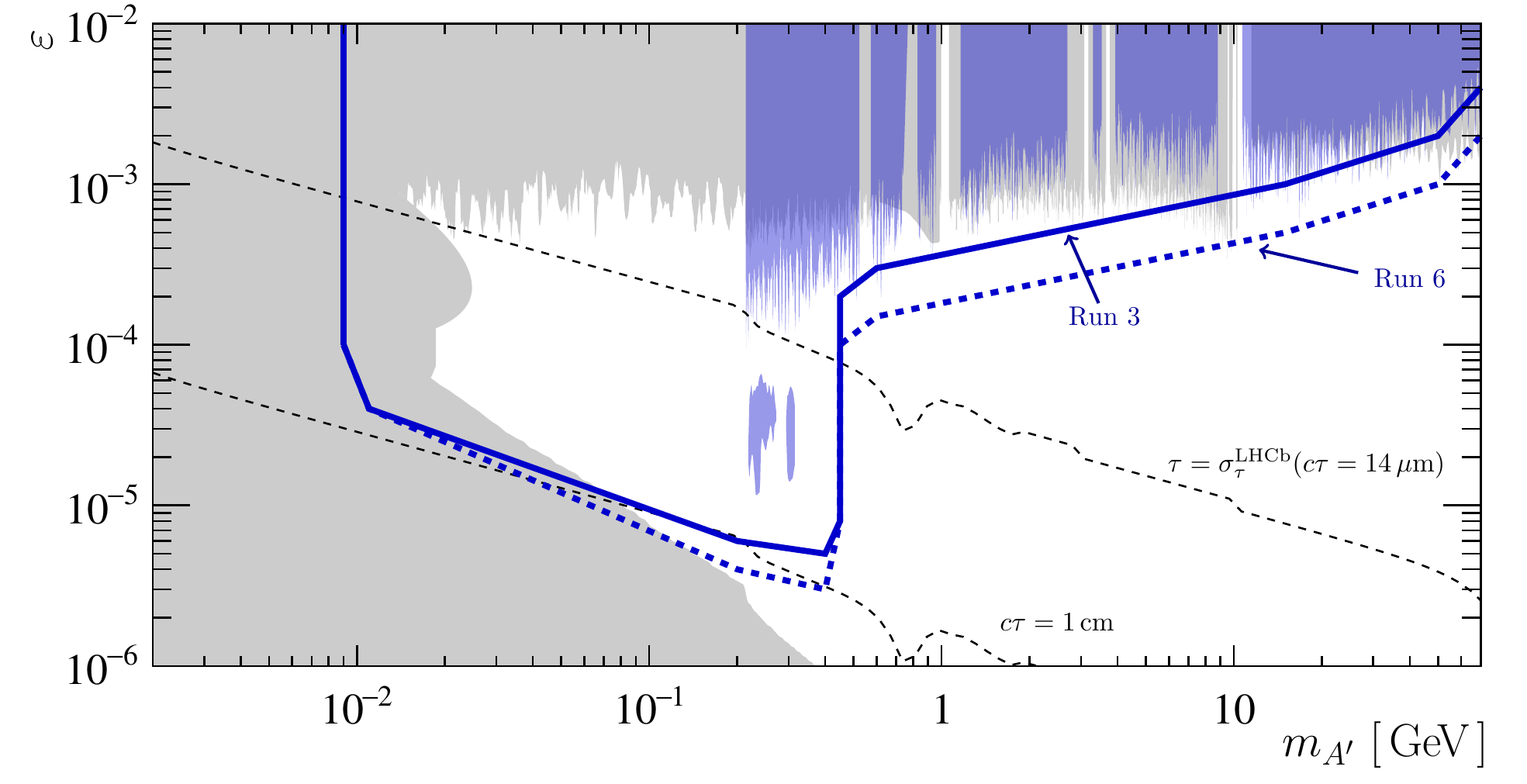}
  \caption{
  Adapted from Ref.~\cite{Graham:2021ggy}:
  constraints on visible $A'$ decays from (blue regions) LHCb~\cite{LHCb-PAPER-2019-031} and (gray regions) all other experiments.
  The solid blue line is the union of Run~3 projections for LHCb from Refs.~\cite{Ilten:2016tkc,Ilten:2015hya}, updated to include inclusive $A' \to e^+e^-$ projections enabled by recent advances in the LHCb trigger. The dashed blue line projects further into the future to the end of Run~6.
  }
  \label{fig:aprime}
\end{figure*}

Reference~\cite{Ilten:2015hya} showed that ultimately the reach of LHCb is limited by the size of its vertex detector (VELO); {\em i.e.}, that the sensitivity is not limited by the signal rate or backgrounds, but instead by the lifetime acceptance. This results in a minimal gain in sensitivity for dark photons going from Run~3 to Run~6, even though the integrated luminosity will increase by a factor of 20.
We can thus trivially estimate the inclusive $A' \to e^+e^-$ sensitivity using the results of Refs.~\cite{Ilten:2015hya,Ilten:2016tkc} by simply connecting the two projections using a line of roughly constant dark-photon lifetime.
The inclusive signal rate will be much larger than the radiative-charm one used in Ref.~\cite{Ilten:2015hya}; however, the background will also be larger, see Refs.~\cite{Ilten:2019xey,CidVidal:2019qub} for signal and background estimates.
These details though only impact the precise luminosity where LHCb sensitivity becomes lifetime limited, not the sensitivity itself at a relevant level.
Figure~\ref{fig:aprime} shows the region we expect LHCb can cover by the end of its operational life.

\section{Higgs portal}

\begin{figure*}[t]
  \centering
  \includegraphics[width=0.99\textwidth]{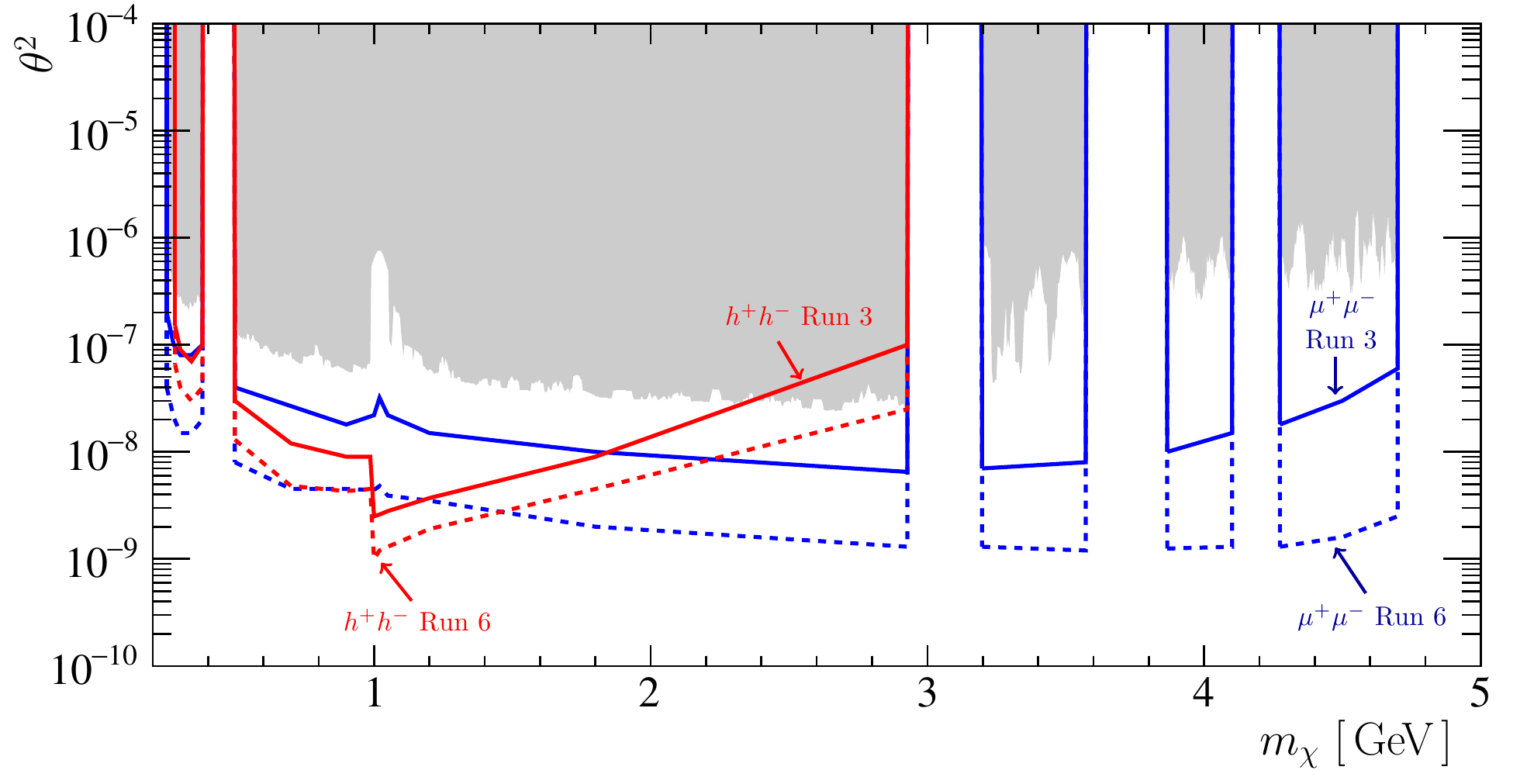}
  \caption{
  Constraints on the Higgs portal from (gray regions) published LHCb searches using Run~1 data~\cite{LHCb-PAPER-2015-036,LHCb-PAPER-2016-052}.
  The solid lines show our projections for Run~3 from (blue) $\chi \to \mu^+\mu^-$ and (red) $\chi \to h^+h^-$, where $h = \pi$ or $K$.
  The dashed lines show the corresponding Run~6 projections. {\em N.b.}, these results assume the quartic coupling is negligible.
  }
  \label{fig:higgs}
\end{figure*}

LHCb published searches for $B \to K^{(*)}\chi(\mu^+\mu^-)$ that placed world-leading constraints on GeV-scale Higgs-portal scalars using Run~1 data~\cite{LHCb-PAPER-2015-036,LHCb-PAPER-2016-052}.
Future projections for the Higgs-portal results were provided in Ref.~\cite{Gligorov:2017nwh}, which are based on the Run~1 results.
Here, we update and improve these projections in the following ways:
\begin{itemize}[leftmargin=1.0em]
    \item Reference~\cite{Gligorov:2017nwh} applied a constant scale factor to predict the future sensitivity based on the Run~1 results; however, this fails to account for the strong lifetime dependence of the LHCb limits. For example, at higher masses the Run~1 data sample only explored prompt $\chi$ decays which are contaminated by SM penguin and charmonium decays. With much greater luminosity, LHCb explores the long-lived $\chi$ region which is background free, improving the sensitivity.
    \item The published LHCb limits are based on older models for the $\chi$ coupling to hadrons. We update the existing limits to use the hadronic couplings in Ref.~\cite{Fradette:2017sdd}, which is the commonly used model employed by recent long-lived-particle experimental proposals. (This update also requires carefully considering the strong lifetime-dependence of the LHCb constraints.)
    \item Finally, we include new projections based on including the hadronic decays $B \to K^{(*)}\chi(\pi^+\pi^-)$ and $B \to K^{(*)}\chi(K^+K^-)$. We show that including these final states can improve the sensitivity in the ${0.5 \lesssim m_{\chi} \lesssim 1.5}$\,GeV region.
\end{itemize}
Figure~\ref{fig:higgs} shows both our updated constraints using the published Run~1 results and our new projections. The Run~1 dimuon searches were background free in the displaced case, and we assume here that this continues to be true throughout the lifetime of LHCb data taking.
Updating the $\chi$ hadronic couplings to those of Ref.~\cite{Fradette:2017sdd} weakens the existing constraints, largely due to the decreased $\chi \to \mu^+\mu^-$ branching fraction.
Figure~\ref{fig:higgs} clearly shows that one cannot use a constant scale factor to predict future sensitivities.
The LHCb dimuon search sensitivity improves more at higher masses. As stated above, going to Run~3 this is largely because the Run~1 data only probed prompt $\chi$ decays, whereas Run~3 will explore the nearly background-free long-lived $\chi$ decays even at higher masses.

For hadronic decays, we only consider the long-lived $\chi$ scenario, as larger couplings are ruled out by the dimuon data alone.
To estimate the background, we consider the related hadronic final state $B^{\pm} \to K^{\mp} \pi^{\pm}\pi^{\pm}$ studied by LHCb in Run~1~\cite{LHCb-PAPER-2016-023}, which is dominated by random combinations of hadrons produced in heavy-flavor decays.\footnote{The motivation here is that we expect the background for $B^{\pm} \to K^{\pm} \chi(h^+h^-)$ for long-lived $\chi$ bosons to largely arise due to combinations of hadrons produced in heavy-flavor decays that randomly satisfy the topological and kinematic constraints, which is also likely the case in our chosen background proxy.}
Assuming that the background uniformly populates each $\chi \to \pi^+\pi^-$ mass window\footnote{This assumption is expected to hold within an order of magnitude, which only affects the limits by up to a factor of two.}---the mass spectra in Ref.~\cite{LHCb-PAPER-2016-023} are integrated over $\pi^+\pi^-$ phase space---we estimate $\mathcal{O}(0.1)$ background candidates per mass window in Run~1.\footnote{We confirm this nearly background-free prediction in Run~1 by noting that $B^{\pm} \to K^{\pm}K_S(\pi^+\pi^-)$ was found to have a background level of only $\approx 20$~\cite{LHCb-PAPER-2013-034}, and since $K_S$ decays have high purity in LHCb, this must be consistent with being entirely random $K_S K$ combinations, {\em i.e.}\ consistent with the non-$K_S$ background (all that is relevant here) being negligible.}
We scale this up for Run~6 accounting for the 100-fold increase in luminosity, along with a roughly factor of two increase in heavy-flavor cross section and a factor of five increase in hadronic-trigger efficiency to obtain an estimate of $\mathcal{O}(100)$ background candidates per mass window.
For concreteness, we take the background to be 100 in Run~6,\footnote{Near $m_D$ the hadronic modes in will in principle be contaminated by a peaking charm background; however, the dimuon search already rules out charm-like lifetimes. Thus, charm backgrounds are unlikely to be a problem and we ignore them here.} and scale this down by the luminosity ratio for Run~3.
The background for $K^+K^-$ is likely lower, but we ignore that here.
Figure~\ref{fig:higgs} shows that including the $\pi^+\pi^-$ and $K^+K^-$ final states improves the sensitivity in the $0.5 \lesssim m_{\chi} \lesssim 1.8$\,GeV region in Run~3, and the  $1.0 \lesssim m_{\chi} \lesssim 1.5$\,GeV region in Run~6.

\section{ALPs coupled to gluons}

LHCb has not yet published searches for ALPs, which are again predominantly produced in penguin decays for GeV-scale masses.
For the case where the dominant ALP-SM coupling is with gluons, the most important decays in the region $0.3 \lesssim m_a \lesssim 3.0$\,GeV are $a \to \eta\pi\pi$ (dominant above 1\,GeV), $a\to \pi^+\pi^-\gamma$, and $a \to 3\pi$~\cite{Aloni:2018vki}.
Therefore, the reactions of interest here are $B \to K^{*}a(\pi^+\pi^-\{\eta,\pi^0,\gamma\})$,\footnote{LHCb can only employ the decays with at least two charged pions. We only use this component of the $\eta\pi\pi$ and $3\pi$ branching fractions from Ref.~\cite{Aloni:2018vki}.}
which will have substantial backgrounds due to the lack of leptons.
Furthermore, the efficiencies for these decays would have been $\mathcal{O}(10)$ times lower than the $B \to K^{(*)}\chi(\mu^+\mu^-)$ decays used in the published Higgs-portal search due to a roughly five-times lower hardware-trigger efficiency and about a factor of two loss in reconstruction efficiency per photon.
However, with the removal of the hardware-trigger stage in Run~3, the efficiency difference will be much smaller moving forward.
Our projections assume that the LHCb efficiency for ALP searches in Runs~3--6 is consistent with its efficiency in Run~1 for the dimuon-based Higgs-portal searches multiplied by $0.5^{n(\gamma)}$, {\em i.e.}\ we take the photon acceptance times efficiency to be 50\%.

Our background estimates for these decays are based on a Run~1 LHCb study of $B^{\pm} \to K^{\pm} \eta'(\pi^+\pi^-\gamma)$~\cite{LHCb-PAPER-2016-060}, where approximately 1400 background candidates were observed in the $\eta'$ mass window.
We scale this up for Run~6 using the same factors as for the hadronic Higgs-portal decays to obtain an estimate of $\mathcal{O}(1{\rm M})$ background candidates per mass window in Run~6.\footnote{Clearly this is a huge background yield per narrow ALP mass window, though we note that the predicted SM $B^{\pm} \to K^{\pm} \eta'(\pi^+\pi^-\gamma)$ yield is 10M. That said, we expect the background will be able to be reduced, but proceed using this conservative estimate.}
Therefore, we take the background to be 1M in Run~6, and scale this down by the luminosity ratio for Run~3.
Reference~\cite{LHCb-PAPER-2016-060} only shows a small region around the $\eta'$, thus we assume no mass dependence in our background estimate.\footnote{We note that our combinatorial estimate is much larger than the $B^{\pm} \to K^{\pm} \pi^+\pi^-\gamma$ yield or any other similar SM yields, which justifies simply ignoring these (which clearly have strong mass dependence). While some mass dependence in the combinatorial background is expected, we doubt this will be orders of magnitude in size, thus not relevant for the level of precision targeted.}
For the $\eta,\pi^0 \to \gamma\gamma$ decays, we simply assume that there is always another photon reconstructed that satisfies the $\eta,\pi^0$ diphoton mass constraint, {\em i.e.}\ we use the same background estimate for these channels even though they require an extra photon where $m_{\gamma\gamma}$ is consistent with the known pseudoscalar-meson masses.\footnote{This is another assumption that will not be precisely true, but should be accurate up to an $\mathcal{O}(1)$ factor.}
Finally, below about 1\,GeV LHCb can eventually explore long-lived ALPs, and even though we expect the background in that scenario to be much lower, we ignore any displacement-based background reduction.

\begin{figure*}[t]
  \centering
  \includegraphics[width=0.99\textwidth]{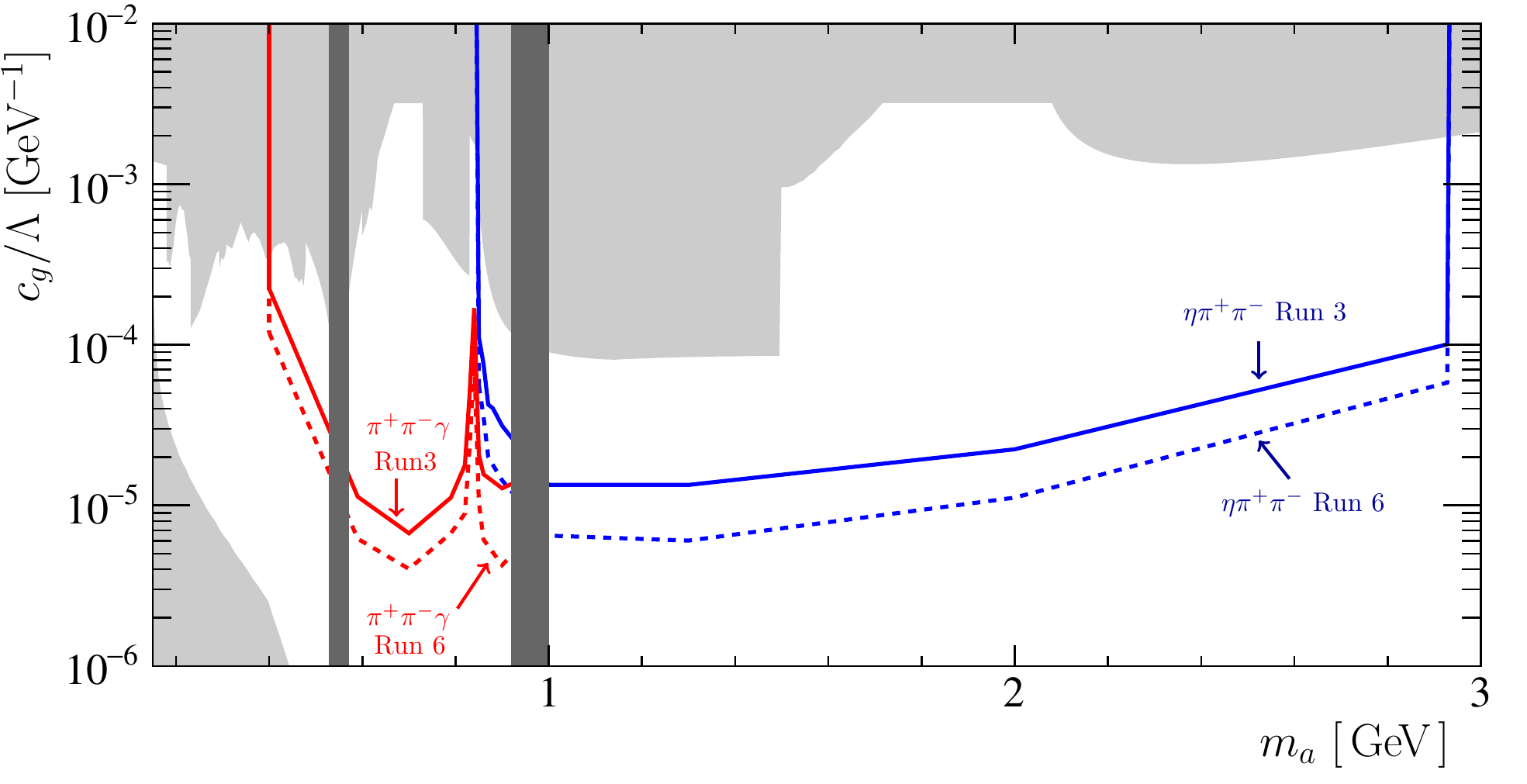}
  \caption{
  Constraints on ALPs from (light gray regions) previous experiments from Ref.~\cite{Aloni:2018vki}.
  The solid lines show our projections for Run~3 from (blue) $a \to \eta\pi^+\pi^-$ and (red) $a \to \pi^+\pi^-\gamma$.
  The dashed lines show the corresponding Run~6 projections. The (dark grey bands) mass regions close to the $\eta^{(\prime)}$ masses are excluded where peaks in the data are expected due to SM processes.
  }
  \label{fig:alps}
\end{figure*}

Figure~\ref{fig:alps} shows the existing constraints from Ref.~\cite{Aloni:2018vki}, along with our projections for LHCb.
We predict that LHCb can greatly improve on existing sensitivity to ALPs that couple to gluons.

\section{Summary}

We provided future LHCb dark-sector sensitivity projections for use in the Snowmass reports.
These include updated projections for dark photons and the Higgs portal, along with new projections for axion-like particles that couple predominantly to gluons.
The Higgs-portal and ALP projections depend strongly on the order of the calculations, and in addition, the ALP results depend on the assumptions made about the unknown UV physics. Once an agreement is made on what assumptions to make to put all projections on equal footing in the Snowmass reports, we will update this document.

\begin{acknowledgements}

This work was supported by NSF grant PHY-1912836.

\end{acknowledgements}

\bibliography{main,LHCb-PAPER,LHCb-DP,LHCb-TDR}

\end{document}